\documentclass[12pt]{article}
\usepackage{times}
\usepackage{geometry}
\usepackage{aas_macros}
\usepackage{color}
\usepackage{comment}
\usepackage[utf8]{inputenc}
\usepackage[utf8]{inputenc}
\usepackage{enumitem,amssymb}
\usepackage{ragged2e}
\newlist{thematic}{itemize}{8}
\usepackage{pifont}
\usepackage{hyperref}
\usepackage{graphicx}

\include{institutionAliases.tex.txt}

\begin{document}
\raggedright
\huge
Astro2020 Science White Paper \linebreak

The Next Generation of Cosmological Measurements with Type Ia Supernovae 
 \linebreak
\normalsize

\noindent \textbf{Thematic Areas:}  Cosmology and Fundamental Physics \linebreak
  
\textbf{Principal Author:}\\
Dan Scolnic \\
Duke University \\
daniel.scolnic@duke.edu\\

\textbf{Contributing Authors:}  Saul Perlmutter (LBNL), Greg Aldering (LBNL), Dillon Brout (Penn), Tamara Davis (Queensland), Alex Filippenko (Berkeley), Ryan Foley (UC Santa Cruz), Ren\'ee Hlo\v{z}ek (Toronto), Rebekah Hounsell (Penn), Saurabh Jha (Rutgers), David Jones (UC Santa Cruz), Pat Kelly (Minnesota), Rick Kessler (U Chicago), Alex Kim (LBNL), David Rubin (STScI), Adam Riess (JHU), Justin Roberts-Pierel (U South Carolina), Steven Rodney (U South Carolina), Christopher Stubbs (Harvard),  Yun Wang (CalTech/IPAC)

\textbf{Endorsed By:} Jacobo Asorey (KASSI),
Arturo Avelino (Harvard),
Chetan Bavdhankar (NCBJ), 
Peter J. Brown (TAMU),
Anthony Challinor (IOA), 
Christophe Balland (LPNHE),
Asantha Cooray (UC Irvine), 
Suhail Dhawan (Oskar Klein), 
Georgios Dimitriadis (UC Santa Cruz), 
Cora Dvorkin (Harvard), 
Julien Guy (LBNL), 
Will Handley (KICC),
Ryan E. Keeley (KASSI), 
Jean-Paul Kneib (EPFL), 
Benjamin L'Huillier (KASSI), 
Massimiliano Lattanzi (INFNFE),
Kaisey Mandel (Cambridge),
James Mertens (York U), 
Mickael Rigault (LPC), 
Pavel Motloch (CITA), 
Suvodip Mukherjee (IAP), 
Gautham Narayan (STScI), 
Andrei Nomerotski (BNL), 
Lyman Page (Princeton), 
Levon Pogosian (Simon Fraser), 
Giuseppe Puglisi (Stanford), 
Marco Raveri (KICP), 
Nicolas Regnault (LPNHE), 
Armin Rest (STScI), 
C\'{e}sar Rojas-Bravo (UC Santa Cruz), 
Masao Sako (Penn), 
Feng Shi (KASSI), 
Srivatsan Sridhar (KASSI), 
Aritoki Suzuki (LBL), 
Yu-Dai Tsai (FNAL), 
W. M. Wood-Vasey (Pitt), 
Yannick Copin (IPNL), 
Gong-Bo Zhao (NAOC), 
Ningfeng Zhu (Penn)


\textbf{Abstract:}

While Type Ia Supernovae (SNe\,Ia) are one of the most mature cosmological probes, the next era promises to be extremely exciting in the number of different ways SNe\,Ia are used to measure various cosmological parameters.  Here we review the experiments in the 2020s that will yield orders of magnitudes more SNe\,Ia, and the new understandings and capabilities to constrain systematic uncertainties at a level to match these statistics.  We then discuss five different cosmological probes with SNe\,Ia: the conventional Hubble diagram for measuring dark energy properties, the distance ladder for measuring the Hubble constant, peculiar velocities and weak lensing for measuring $\sigma_8$, and strong-lens measurements of H$_0$ and other cosmological parameters.  For each of these probes, we discuss the experiments that will provide the best measurements and also the SN\,Ia-related systematics that affect each one.

\pagenumbering{gobble}
\newpage
\pagenumbering{arabic}

\section{Introduction}

~~~~~SNe\,Ia were used to discover the current epoch of acceleration of the universe with less than 100 objects \cite{Riess98,Perlmutter99}.  They remain one of the premier probes of cosmology owing to their high luminosity, excellent standardizability, and relatively high rates that allow SNe\,Ia to be used to probe multiple epochs of the universe.  In the 2020s, surveys will discover hundreds of thousands of SNe\,Ia across a large redshift range ($0<z<2$), and together with our advances in understanding and controlling the sources of systematic uncertainties, this opens up a wide array of potential uses.  The common mode for leveraging distances of SNe\,Ia to measure cosmological parameters is to compare SN\,Ia distances in different redshift regimes: to measure dark energy properties, SN\,Ia distances at $z\approx0.05$ and $z\approx0.5$--1.5 are compared; to measure the Hubble constant, SN\,Ia distances at $z\approx0.005$ and $z\approx0.05$ are compared. These are probes to measure the geometry of the universe. However, with a large boost in statistics, we can also measure the growth of structure of the universe by searching for additional information that can be found in the scatter about the best-fit cosmological model.  At low-$z$, this yields constraints on the model of gravity via peculiar velocities.  At high-$z$, this yields information into the magnitude of weak lensing of the SNe\,Ia.  Furthermore, with increased statistics, there is a higher likelihood of discovering strongly lensed SNe\,Ia which can provide a direct measurement of H$_0$.

~~~~~There are typically two modes for capturing information about SNe\,Ia.  The first is imaging;  this is good for light curves, but if used by itself, is dependent on the success rate of photometric classification of SNe\,Ia and redshift determination.  The second mode is spectroscopy;  this is good for spectroscopic redshifts of host galaxies, classification of the SNe, and feature extraction. Pixelized, photometric spectroscopy (spectrophotometry) can further make possible fine distinctions among SNe\,Ia that can allow more precise distance measurements and control of systematics from evolutionary population drift and dust \cite{Chotard11, Fakhouri15,Rigault13,Nordin18}.

\begin{figure}
\begin{center}
\includegraphics[width=\textwidth]{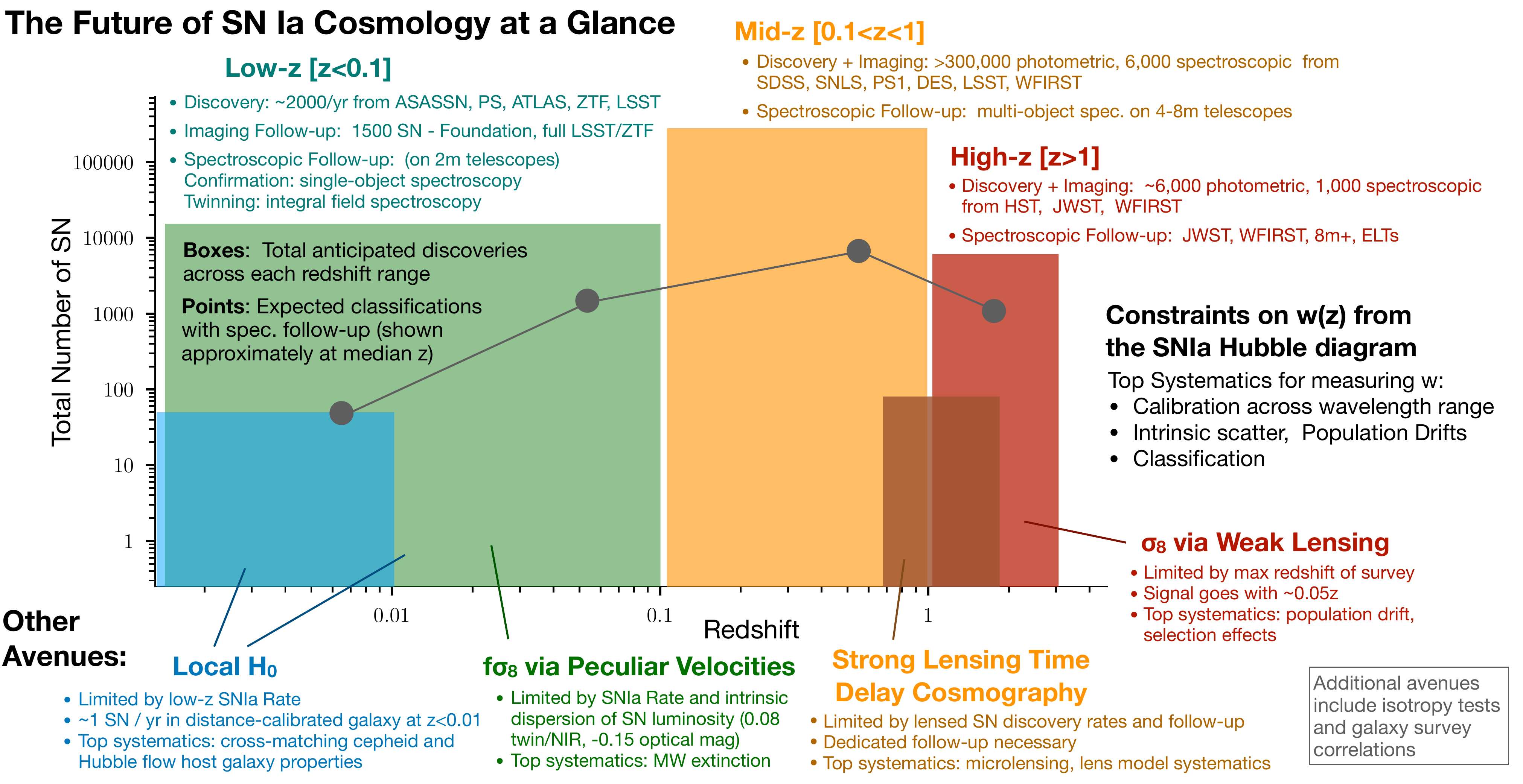}
\caption{Summary of cosmological tests using Type Ia SN in the 2020's.} 
\label{imaging}
\end{center}
\end{figure}

~~~~~To achieve these next-generation science goals with such SN measurements, a panoply of telescopes will be needed, with the appropriate instrumentation. This is summarized in Figure~1, which details the important telescopes for each redshift regime.  In the text below, we expand on this figure and describe the five different cosmological probes that use SNe\,Ia.  The general structure is to first outline the approach, second describe samples that have been or could be used, and third describe the systematics.

\newcommand{\method}[5]{
\noindent \textbf{#1:} \\
\underline{Description:} #2 
\underline{Experiments:} #3
\underline{Approach:} #4
\underline{Key Systematics:} #5
}

\section{The Conventional Hubble Diagram}
\label{sec:2}

\textbf{Approach:} The main approach is to fit cosmological models to the SN distance versus redshift (the ``Hubble diagram"). Efforts are focused on distinguishing between models of dark energy, by continually increasing the statistics and redshift range, while improving our techniques to measure and control the systematic uncertainties to match these improved statistical uncertainties.  

\textbf{Samples:} The data used to constrain cosmological models can be combined from multiple samples and surveys.  Recent analyses (e.g., \cite{Scolnic18}) have included data from 10$+$ SN\,Ia samples, including \cite{Riess98,Jha06,Hicken09b,Amanullah10,Hicken12,Stritzinger11,Rest14,Astier06,Guy10,Betoule14,Holtzman08,Riess04,Riess17,Suzuki12}.  Cosmologically viable SN\,Ia redshifts range from 0.01 to $\sim2.3$, where the minimum redshift is chosen because of limitations in peculiar-velocity estimates and the maximum redshift is that of the farthest SNe\,Ia studied \cite{Jones15, Rubin13}.  Most cosmological analyses have focused on spectroscopically-confirmed samples; but recently, larger photometric samples \cite{Campbell13,Jones17,Jones18} have been considered.   The boost in statistics from ongoing/recent surveys will be on the order of $5\times$ at low-$z$ (owing to ZTF, Foundation, ATLAS, ASAS-SN \cite{ztf, Foley17, ATLAS, ASASSN} and $2\times$ at mid-$z$ (owing to DES).  The boost in statistics from future surveys will be on the order of $300\times$ at mid-$z$ (owing to LSST \cite{LSSTSRD}) and $1000\times$ at high-$z$ (owing to WFIRST \cite{Hounsell18,Spergel15}).  JWST will be able to extend to even higher redshifts, perhaps $z \approx 5$ --- if the SN\,Ia rate is sufficiently high at that epoch for discovery.  A next-generation space mission such as LUVOIR or perhaps OST would further extend the Hubble diagram.

\textbf{Systematics:} For much of the last decade, the largest systematic uncertainties have been related to photometric calibration (e.g., \cite{Betoule14}).  The systematics related to photometric calibration are the result of uncertainty in individual zeropoints \cite{Hicken09a}, cross-survey calibration \cite{Supercal}, proper calibration of the model SN\,Ia SED \cite{Betoule14,Saunders15,Saunders18}, and HST-based calibration \cite{Bohlin15,Rubin15a}. As significant effort has been made on calibration systematics, recent papers (e.g., \cite{Scolnic18}, \cite{Brout18SYS}) have shown that the uncertainties in the intrinsic scatter model of SN\,Ia brightnesses and the underlying color populations can have similar contributions to the systematic error budget as does calibration, and there will be additional systematics owing to photometric classification \cite{Jones17} and host-galaxy misassociation \cite{Gupta16} that will be significant in future studies using photometric samples. The systematic control for future surveys like WFIRST and LSST is on the mmag level, and the sensitivity to specific systematics is discussed by \cite{LSSTSRD} and \cite{Hounsell18}.  Additional systematics related to SN\,Ia evolution, dust properties, and other astrophysical factors can be addressed by both larger samples to discover correlations and high-quality (wide-wavelength-range, host-subtracted) spectrophotometry of high-redshift SNe\,Ia.  Photometric classification errors and K-correction errors can be reduced by producing large spectroscopic samples of all classes of SNe through redshifts beyond $z =1$; this will require that the appropriate ground- and space-based spectroscopic instrumentation be available during the SN discovery surveys.  Spectroscopy of hosts, using wide-field multi-object spectrographs, will further enhance the photometric samples.

\section{The Local Distance Ladder}
\label{sec:3}

\textbf{Approach:} SNe\,Ia are used in two out of the three rungs in the distance ladder, as presented by \cite{Riess16} and \cite{Freedman12}.  SNe\,Ia in the second rung of the distance ladder, in which Cepheid variables are used to calibrate SN\,Ia distance measurements, are denoted as the calibrator set.  SNe\,Ia in the third rung are denoted as the Hubble-flow set.  The mean offset in brightness between these two sets yields the Hubble intercept, which can be used independently of the Cepheid measurements in the final H$_0$ calculation (see, e.g., \cite{Burns18}).  After upcoming improvements to the geometric calibration of Cepheids and other stellar types from Gaia and detached eclipsing binaries in the LMC, we can expect the measurement of H$_0$ to be limited
by the sample size of SNe\,Ia in the hosts of these stellar calibrators. 

\textbf{Samples:} The use of SNe\,Ia in the Hubble-flow sample is described in the above section.  Typically, a maximum redshift cut is applied (at $z=0.15$ as by \cite{Riess16}, or $z=0.5$ as by \cite{Kenworthy19}), depending on how the nonlinear terms are calculated.  The minimum cut has typically been set at $z=0.023$ owing to legacy concerns of a ``Hubble bubble" but can be reduced to $z=0.01$.  The second rung of the distance ladder utilizes a ``calibrator" sample of nearby SNe\,Ia ($\lesssim 40$ Mpc or $z \lesssim 0.01)$ discovered in galaxies also hosting Cepheid variables.  \cite{Riess16} utilize 19 SNe\,Ia in nearby galaxies, observed by a variety of telescopes over the last four decades.    Efforts to improve the per-SN-precision (for high-quality data) and growing this sample are likely to reach a floor
of $10^2$ SNe\,Ia, potentially with the help of future telescopes like JWST or LUVOIR, and a best precision of 5\% per object, resulting in a $\sim0.5$\% limit.  With additional efforts to match the calibrator and Hubble-flow sample ($\sim0.5$\% limit), it is not unrealistic to presume a $\sim1$\% measurement of H$_0$ via calibrated SNe\,Ia.  This precision is crucial to approach that available from the CMB to fully exploit this test of the cosmological model. 

\textbf{Systematics:} The key systematic uncertainties are those that affect the calibrator and Hubble-flow sets differently.  While for the Hubble diagram, the key systematic uncertainties are due to differences in calibration or physics between high-$z$ and low-$z$ datasets, for H$_0$ these types of systematics are largely mitigated because the photometric systems and selection criteria are similar and the change in redshift is limited to $\Delta z \approx 0.05$.  There is potential sensitivity to systematics related to the host-galaxy environment of the SNe\,Ia as the calibrator set includes only star-forming host galaxies \cite{Rigault15,Rigault18}, and while recent analyses find the impact for current H$_0$ measurements is $<1$\% (see \cite{Jones18}; \cite{Rose19}), work on this topic is ongoing and future measurements can be less sensitive to this systematic by curating a Hubble-flow sample to be more representative of the calibrator sample.

\section{Measuring Gravity with Peculiar Velocities -- $f\sigma_8$}
\label{sec:4}

\textbf{Approach:} Peculiar velocities (PVs) refer to galaxy motion that deviates from the homogeneous expansion of the universe.  They arise owing to the gravitational pull from large-scale structure in the universe.  Therefore, they have the potential of directly probing the total distribution of matter, including dark matter, in contrast to measuring the distribution of just the galaxies.   PVs are measured by comparing the Hubble residuals of SN\,Ia distances.  The accelerating expansion of the universe could be driven by non-general-relativity (GR) gravity, and measuring the peculiar velocities measures the gravity model. This is parameterized by the growth rate of structure, $f \approx \Omega_M^\gamma$. The local PV power spectrum measures $f{\sigma_8}$ and comparison with high-$z$ redshift-space distortions (RSDs; e.g., from DESI or 4MOST) measures $\gamma$ and thus deviations from GR. At low redshift, direct measurement of PVs using SNe\,Ia overcomes the cosmic variance floor experienced by statistical methods (e.g., RSD) for measuring $f{\sigma_8}$.


\textbf{Sample:}  The best current constraint from SNe\,Ia is $f{\sigma_8}=0.428\pm0.0465$ for $z_{{eff}}=0.02$ \cite{Huterer17} with a low-redshift SN sample of $\sim300$. Since the SN\,Ia density is too low to sample all modes in a volume, the variance on $f{\sigma_8}$ improves linearly with sample size. \cite{Howlett17} discusses how measurement of $f{\sigma_8}$ can be improved with LSST with up to 18k or 160k SN\,Ia light curves, depending on assumptions about the level of LSST SN classification and redshift determination.  
The weight of each SN increases as the square of precision of SN velocity measurements. At low-$z$, where most of the leverage resides, the sample size is limited by the rate of SNe\,Ia.  Typical distance precision from photometric surveys is $0.15$~mag, or $0.07 c z$; subtyping using wide-wavelength-range, host-subtracted spectrophotometry obtained at maximum light reduces this uncertainty to $0.035 c z$ \cite{Fakhouri15}. 
NIR coverage also shows promise for improving SN\,Ia PV measurements.
Aggressive follow-up observations of the large SN\,Ia samples now becoming available could reduce the uncertainty in $f{\sigma_8}$ at low redshift to a few percent.

\textbf{Systematics:} This approach is less sensitive to systematics than other probes because of the smaller impact of
color-calibration uncertainties and SN population evolution in the restricted redshift range. 
Uncorrected power-spectrum modes on scales outside the survey volume set the error floor.
Cross-survey calibration
and Milky Way extinction are important systematics because they project onto the power spectrum in the angular direction. 
SN standardization residuals correlated with host galaxies \cite{Rigault13,Rigault18} could also introduce systematics since galaxy properties correlate with structure in the density field.

\section{Measuring The Weak-Lensing Signal}
\label{sec:5}

{\bf Approach:} SNe\,Ia are lensed by structure in the Universe, which amplifies or deamplifies their brightness, depending on whether the line-of-sight to the SN passes through overdense or underdense regions \cite{Holz98,Holz05}.  With sufficient statistics and proper modeling, weak-lensing magnification can measure cosmological parameters either using the magnification distribution of SNe\,Ia \cite{Wang99,Wang05}, by cross-correlating magnitudes with the density observed along the line of sight, or by using amplification to assess dark-matter halo properties \cite{Metcalf99,Goliath00,Jonsson08}.  

{\bf Sample:}  Lensing magnification of SNe\,Ia was marginally detected by \cite{Smith14} who cross-correlated SN\,Ia magnitudes from the SDSS sample with the density observed along the line of sight.  \cite{Zhai18} have reconstructed the lensing distribution $p(\mu)$ using current data.  An ongoing follow-up to this study finds that $\sim600$ SNe\,Ia per redshift bin are required to accurately reconstruct $p(\mu)$ at the mean redshift. Such data at $z>1$ can be obtained by measuring host-galaxy redshifts for all the SNe\,Ia discovered by WFIRST. This will yield a powerful SN\,Ia dataset that can be used to reconstruct $p(\mu,z)$ in multiple redshift bins at $1<z<2$, which will provide an independent probe of cosmology in that redshift range. 

{\bf Systematics:} Theoretical uncertainty in the expected lensing distribution will be one of the largest systematics, as the distribution is sensitive to both large- and small-scale structure that is difficult to model.  In addition, it will be important to understand pollution from misclassification of SNe\,Ia, intrinsic scatter of SNe\,Ia, possible non-Gaussianity in the uncertainty of SN\,Ia magnitudes, and how these interplay with selection effects.

\section{Strongly Lensed SNe\,Ia}
\label{sec:5}

{\bf Approach:} 
The method of {\it time-delay cosmography} 
involves measuring the delay in arrival time between each 
of the multiple images of a strongly lensed transient. Combining 
this with a model for the lensing potential, one can derive 
a ratio of cosmological distances to the source and the lens 
\cite{Refsdal64}.  This distance ratio is inversely 
proportional to H$_0$, and weakly sensitive to other 
cosmological parameters.  
Time-delay cosmography has been successfully employed for decades
using multiply-imaged {\it quasars} \cite{Treu16, Suyu18}. Gravitationally lensed 
SNe with multiple images can provide a valuable new addition to the field.

{\bf Sample:}  The first multiply-imaged Type II
(Refsdal: \cite{Kelly16})) and Type Ia ( iPTF16geu: \cite{Goobar17}) SNe have been discovered in the past few years.   LSST is expected to 
deliver hundreds to thousands of strongly lensed transients --- though 
extensive follow-up observations will be needed to make them useful for time-delay cosmography
(\cite{Oguri10}; \cite{Goldstein18}).  This 
will be complemented by a high-$z$ sample from WFIRST, which 
by itself could deliver at least several dozen well-sampled lensed SNe 
and quasars at $z>1$ \cite{Oguri10}.   Linder (2011) 
worked out a baseline expectation for the impact of such a 
sample, assuming 150 lensed quasars and SNe from LSST and WFIRST, each 
with an (admittedly optimistic) 1\% time-delay measurement.  
Combining this sample with information from future SN\,Ia and 
CMB experiments could improve the figure of 
merit for time-variable dark energy parameters ($w_0$, $w_a$)
by a factor of almost 5, in part because time-delay cosmography 
is naturally complementary to SN\,Ia and CMB cosmological 
constraints.

{\bf Systematics:} The major limitations on the accuracy of 
time-delay cosmography have been 
(1) systematic uncertainties in measurement of time delays, particularly due to microlensing,
(2) systematic uncertainties associated with modeling the primary deflector, and 
(3) accounting for lensing effects from other mass along the line of sight.
For item 1, systematic errors in lensed SN time-delay measurements
may reach $\sim4$\% \cite{Goldstein18}.
Relative to lensed quasars, microlensing uncertainties should be easier to characterize
in lensed SNe because they have simpler and more predictable light curves.
Furthermore, when a luminosity distance
to a lensed SN can be determined (e.g., for a lensed SN\,Ia), this
can provide important leverage for breaking lens model degeneracies (items 2 and 3).

\section{More Cosmological Probes with SN\,Ia, and Conclusions}

~~~~~Other SNe\,Ia-based cosmological probes, beyond these five, are also being studied:  e.g., measurements of anisotropy \cite{Soltis19,Brownsberger19} that can use the thousands of SNe\,Ia from LSST and WFIRST;  and a multi-tracer generalization of the Alcock-Paczynski test that can cross-correlate SN\,Ia  with galaxy-redshift surveys \cite{Wandelt19}. With the now obtainable statistical and systematic uncertainties, the 2020s promise many exciting  uses of SNe\,Ia to better understand the universe.

\pagebreak

\bibliographystyle{unsrt}

\end{document}